\documentclass[a4paper]{jpconf}
\pdfoutput=1
\usepackage{graphicx}
\usepackage{mathrsfs}
\usepackage{amssymb,amsmath, amsthm}
\usepackage{xcolor}
\usepackage{slashed}
\usepackage{color}

\begin{document}

\title{Quark-antiquark antenna splitting in a medium}

\author{Fabio Dom\'inguez, Carlos A Salgado and V\'ictor Vila} 

\address {Instituto Galego de F\'isica de Altas Enerx\'ias (IGFAE), Universidade de Santiago de Compostela, E-15706 Santiago de Compostela, Galicia, Spain }

\ead {victor.vila@usc.es}

\begin{abstract}
We study a system in which a quark-antiquark antenna emits a hard gluon in a medium, and an extra very soft emission afterwards. Considering the coherence effects in terms of the survival probability, we can describe the interaction of the $ q\bar{q}g $ configuration with the medium. We extrapolate previous analyzes of the antenna radiation to the case of two hard splittings inside the medium, and prove that this generalization keeps back the picture of jet quenching with effective emitters in the parton QCD cascade.
\end{abstract}

\section{Jet substructure}
The study of jets has evolved into one of the main activities attempting to prove the quark-gluon plasma properties in heavy ion collisions at the LHC. In spite of lacking a rigorous theoretical framework for jet evolution in matter, exhaustive measurements of jet substructure shed light on the character of the jet interactions with the dense medium formed in the collisions. In order to acquire a consistent understanding of the underlying components responsible for the observed results, several questions remain to be answered. For example, is energy loss predominantly caused by medium induced radiation?, do partons lose energy coherently?, or how does the medium response to the jet propagation? 

\section{$ q\bar{q} $ antenna as a laboratory}
The concept of color coherence is very effective to understand the aspects of the QCD radiation, see e.g. Ref. \cite{Dokshitzer:1991wu}. In a parton cascade, when a highly collinear parton is emitted during the propagation, it is difficult to resolve the individual radiated partons, and for large angles their radiation interfere rather than add incoherently - which means that their net behaviour is quite close to the radiation from the parent parton. Subsequent parton branchings of the shower occur in a coherent  manner.

The singlet antenna spectrum for $ q \bar{q} $ production plus soft gluon emission is
\begin{equation}
dN=\frac{d\omega}{\omega}\frac{d\Omega}{2\pi}\frac{\alpha_{s}C_{F}}{2 \pi}\Big[R_{q}+R_{\bar{q}}- 2\mathcal{J}\Big],
\end{equation}
where $ R_{q} $ ($ R_{\bar{q}} $) is the radiation spectrum off an independent constituent, and $ \mathcal{J} $ describes the quark-antiquark interference ($ \omega $ is the energy of emitted gluons, $ d\Omega $ is the differential solid angle, $ \alpha_{s} $ is the coupling constant in QCD and $ C_{F} $\footnote{$ C_{F}=\frac{N_{c}^{2}-1}{2N_{c}} $, with $ N_{c} $ the number of colors.} is a color factor).

This spectrum is divergent in the soft and collinear gluon emission limits. The spectrum is also suppressed at large angles because of the appearance of destructive interferences (just when $ \mathcal{R}_{q}+\mathcal{R}_{\bar{q}} \simeq 2\mathcal{J} $). Therefore, large-angle gluon emission in vacuum is sensitive to the total charge of the system, which leads to angular ordering of the vacuum cascade.

\section{In-medium antenna radiation}
The main contribution to the medium-induced radiation for the case of a very soft gluon emission taking place outside the medium can be easily computed from the diagrams of Fig. 1. Here the resulting spectrum of emitted gluons is a modification of eq. (1) \cite{MehtarTani:2011tz}

\begin{equation}
dN=\frac{d\omega}{\omega}\frac{d\Omega}{2\pi}\frac{\alpha_{s}C_{F}}{2 \pi}\Big[R_{q}+R_{\bar{q}}-(1-\Delta_{med}) \ 2\mathcal{J}\Big].
\end{equation}

\noindent
We introduce the survival probability

\begin{equation}
\mathcal{S} \equiv 1-\Delta_{med} \equiv \frac{1}{N_{c}^{2}-1} \ Tr \ \Big<W(\vec{x}_{\perp})W^{\dagger}(\vec{y}_{\perp})\Big>,
\end{equation}
which describes the interaction of the $ q \bar{q} $ pair with the medium, and the Wilson lines describe the propagation of a gluon through a medium field $ \mathcal{A}_{-}(x_{+},\vec{x}) $,

\begin{equation}
W(\vec{x})=\mathcal{P} \ \exp \Big[ig \int dx_{+} \mathcal{A}_{-}(x_{+},\vec{x}) \Big].
\end{equation}

\parindent=0mm
We have also introduced the decoherence parameter, $ \Delta_{med} $, a factor which determines a characteristic time scale for decoherence of the $ q\bar{q} $ pair. For the case of a static medium of size $ L $, it results to be 

\begin{equation}
\Delta_{med} \simeq 1-\exp \Big[-\frac{1}{4} \hat{q} L (\vec{x}_{\perp}-\vec{\bar{x}}_{\perp})^{2} \Big],
\end{equation}
where $ \hat{q} $ is the transport coefficient. This is the main parameter to be determined by fits and to be compare with theoretical calculations - it encodes all the information about the medium properties.

\parindent=5mm
Here we can study two different limits \cite{CasalderreySolana:2012ef,MehtarTani:2011tz}

\begin{enumerate}
\item when the color correlation length of the medium ($ \simeq \frac{1}{\sqrt{\hat{q}L}} $) is larger than the size of the pairs, $ \Delta_{med} \rightarrow 0 $, the medium cannot resolve the individual emitters, thus acting like an individual object with the total charge of the pair ($ dN \simeq R_{q}+R_{\bar{q}}-2\mathcal{J} $).

\item in contrary case, when $ \Delta_{med} \rightarrow 1 $, the medium resolves the antenna and it breaks the color coherence of the pair, behaving as two independent emitters ($ dN \simeq R_{q}+R_{\bar{q}} $).
\end{enumerate}

\begin{figure}
\begin{center}
\includegraphics[width=0.5\textwidth]{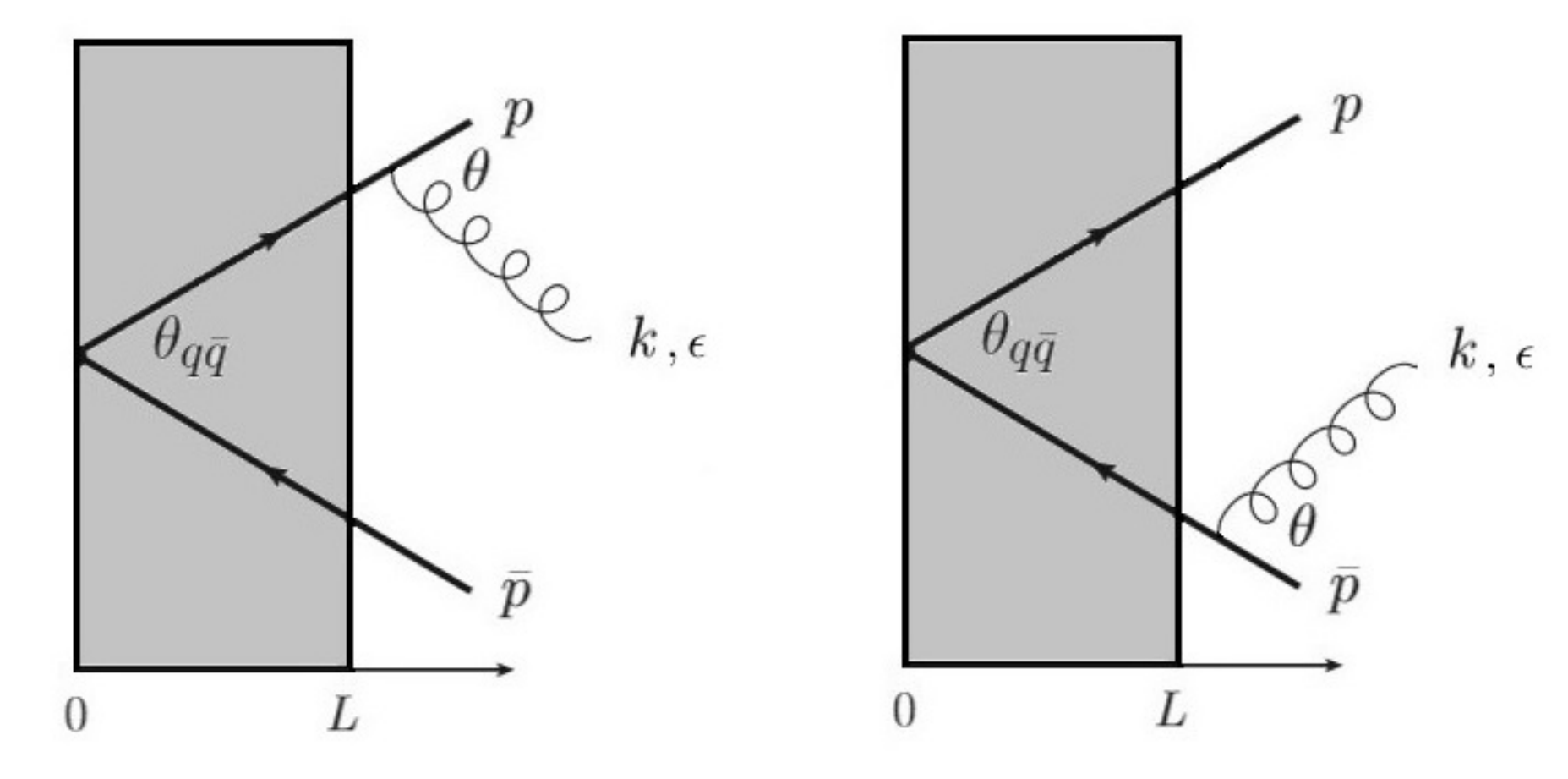}
\end{center}
\label{fig:antennarad}
\caption{Out-out radiation for a very soft gluon emission outside the medium ($ \omega \rightarrow 0 $).}
\end{figure} 

\section{Multiple emissions}
The previous computations are subject to many approaches to simplify the problem. Indeed, one of the main limitations of the original antenna laboratory is that it only includes one in medium splitting. Now, we raise a multiple emissions problem to gloss over this limitation. Particularly, we repeated the previous calculations including an additional hard gluon, with the final goal of extrapolating our results to n-gluon emissions. We restrict to the case of very soft gluon radiation ($ \omega \rightarrow 0 $) [4].

Let us consider now the case of two hard splittings inside the medium. The direct terms $ R_{q}, R_{\bar{q}} $ and $ R_{g} $ are proportional to a color factor, i.e., no medium effects appear as expected with the kinematics chosen.

\newpage

\begin{figure}[h!]
\begin{center}
\includegraphics[width=0.9\textwidth]{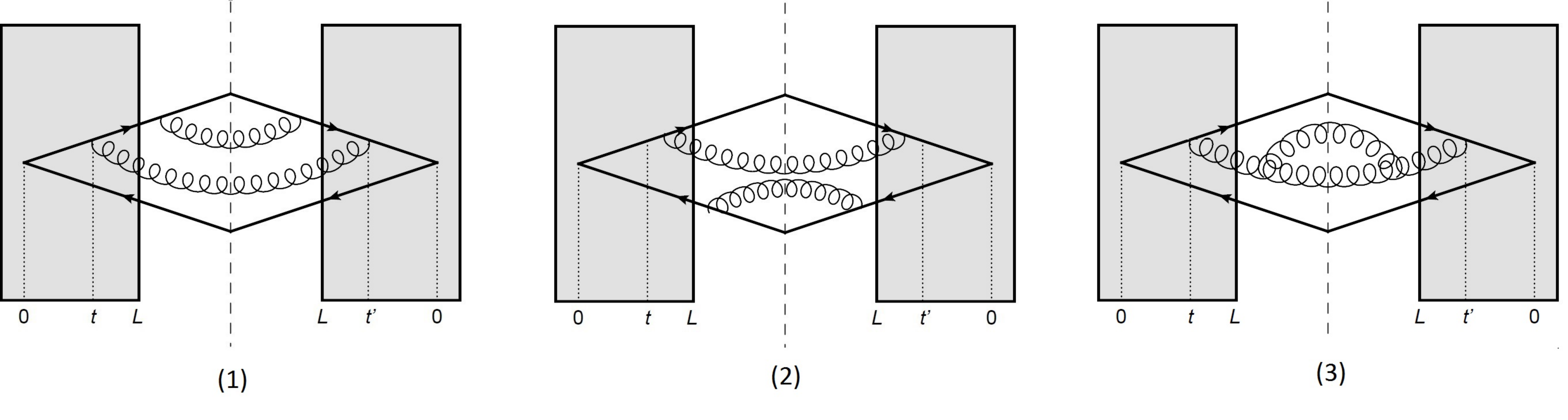}
\end{center}
\caption{Direct terms for the case of three emitters.}
\end{figure} 

\begin{equation}
|\mathcal{M}_{1}|^{2} \propto C_{F}^{2} \ \ \ \ \ \ \ \ \ \ \ \ \ \ \ \ \ \ \ \ \ |\mathcal{M}_{2}|^{2} \propto C_{F}^{2} \ \ \ \ \ \ \ \ \ \ \ \ \ \ \ \ |\mathcal{M}_{3}|^{2} \propto N_{c} \ C_{F}^{2}
\end{equation}

\

In the interference terms, we obtain interesting results because of the explicit presence of the survival probabilites $ \mathcal{S} $, providing information about the color coherence of the emitters.

\begin{figure}[h!]
\begin{center}
\includegraphics[width=0.9\textwidth]{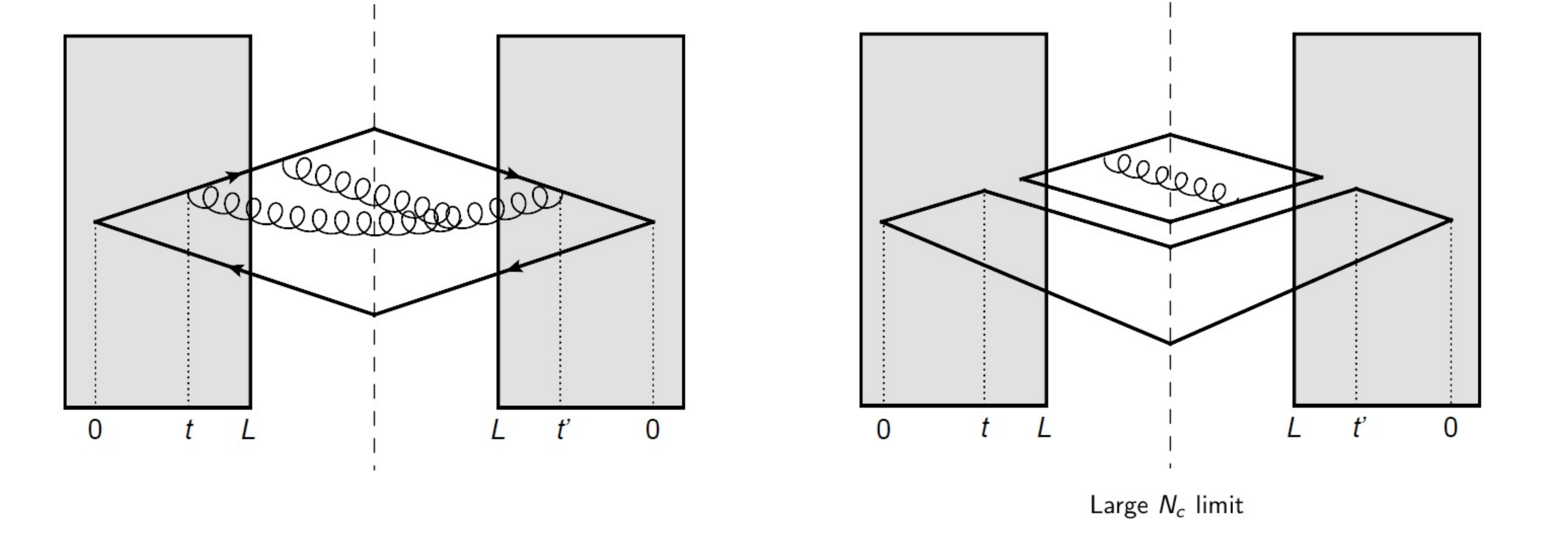}
\end{center}
\caption{Interference term for the case of three emitters (soft gluon in amplitude).}
\end{figure} 

\begin{equation}
\mathcal{M}_{1} \otimes \mathcal{M}_{3}^{*} \propto \mathcal{S}(t,L) \label{10}
\end{equation}

\begin{figure}[h!]
\begin{center}
\includegraphics[width=0.9\textwidth]{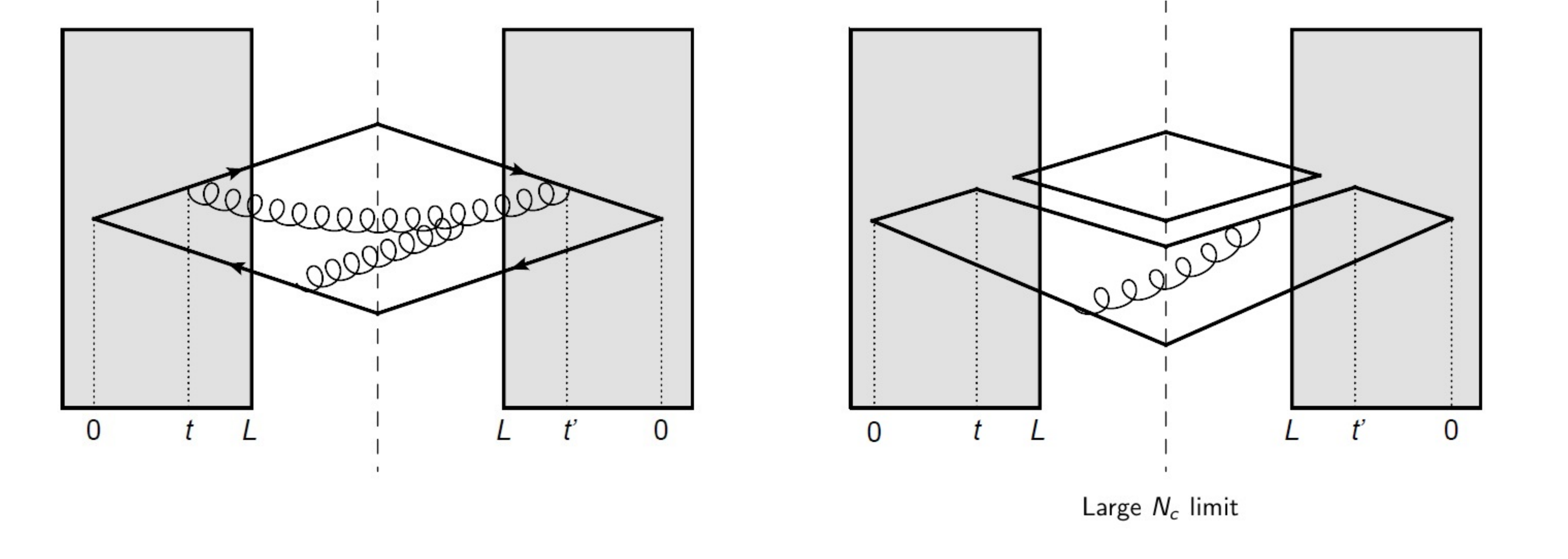}
\end{center}
\caption{Interference term for the case of three emitters (soft gluon in conjugate amplitude).}
\end{figure} 

\begin{equation}
\mathcal{M}_{2} \otimes \mathcal{M}_{3}^{*} \propto \mathcal{S}(0,t) \ \mathcal{S}(t,L) \label{11}
\end{equation}

\

The interpretation of our results is very clear in the large-$ N_{c} $ limit (see Figs. 3 and 4). The survival probability, controlling the degree of color coherence in the soft gluon emission, corresponds to that of the dipole from which this gluon is emitted. In Fig. 3 this dipole is formed at time $ t $ during the evolution of the system, while it is produced at time $ 0 $ in the diagrams of Fig. 4. In this second case, the survival probability is the product of the two survival probabilities, from $ 0 $ to $ t $ and from $ t $ to $ L $. The conclusion is that the general result of the antenna is valid for each of the smaller antennas.

\section{Discussion and conclusions}
The color coherence phenomenon is basic to understand the jet constituents' energy loss. In spite of the singlet antenna limitations (eikonal propagation, only one splitting in medium, zero formation time...), it is a very convenient laboratory. In the multiple emissions setup, the general result of the antenna can be extrapolated for each of the smaller antennas. Finally, these computations go a step forward to obtain a complete description of a QCD shower.

\ack
This research was supported by the European Research Council grant HotLHC ERC-2011- StG-279579, Ministerio de Ciencia e Innovaci\'on of Spain under project FPA2014-58293-C2-1-P and Xunta de Galicia (Conseller\'ia de Educaci\'on) - the group is part of the Strategic Unit AGRUP2015/11 and IGFAE is a Maria de Maetzu Unit of Excellence.

\end{document}